# A Deadline and Budget Constrained Cost-Time Optimisation Algorithm for Scheduling Task Farming Applications on Global Grids

Rajkumar Buyya[†], Manzur Murshed[*], and David Abramson[†]

[†] School of Computer Science and Software Eng.
Monash University, Caulfield Campus
Melbourne, Vic 3145, Australia
*rajkumar@buyya.com, davida@csse.monash.edu.au*

[*] Gippsland School of Computing and IT
Monash University, Gippsland Campus
Churchill, Vic 3842, Australia
*Manzur.Murshed@infotech.monash.edu.au*

**Abstract:** Computational Grids and *peer-to-peer* (P2P) networks enable the sharing, selection, and aggregation of geographically distributed resources for solving large-scale problems in science, engineering, and commerce. The management and composition of resources and services for scheduling applications, however, becomes a complex undertaking. We have proposed a computational economy framework for regulating the supply and demand for resources and allocating them for applications based on the users' quality of services requirements. The framework requires economy driven *deadline and budget constrained* (DBC) scheduling algorithms for allocating resources to application jobs in such a way that the users' requirements are met. In this paper, we propose a new scheduling algorithm, called *DBC cost-time optimisation*, which extends the DBC cost-optimisation algorithm to optimise for time, keeping the cost of computation at the minimum. The superiority of this new scheduling algorithm, in achieving lower job completion time, is demonstrated by simulating the World-Wide Grid and scheduling task-farming applications for different deadline and budget scenarios using both this new and the cost optimisation scheduling algorithms.

## 1 Introduction

Computational Grids [1] and *peer-to-peer* (P2P) computing [2] networks are emerging as next generation computing platforms for solving large-scale computational and data intensive problems in science, engineering, and commerce. They enable the *sharing*, *selection* and *aggregation* of a wide variety of geographically distributed resources including supercomputers, storage systems, databases, data sources, and specialized devices owned by different organizations. However, resource management and application scheduling is a complex undertaking due to large-scale heterogeneity present in resources, management policies, users, and applications requirements in these environments.

The resources are heterogeneous in terms of their architecture, power, configuration, and availability. They are owned and managed by different organizations with different access policies and cost models that vary with time, users, and priorities. Different applications have different computational models that vary with the nature of the problem. The *resource owners* and *consumers*/*end-users* have different goals, objectives, strategies, and demand patterns. In our earlier work [4]–[8], we investigated the use of economics as a metaphor for management of resources in Grid computing environments. The computational economy framework provides a mechanism for regulating the demand and supply for resources and allocating them to applications based on the users' quality of services requirements. It also offers incentive to resource owners for sharing resources on the Grid and end-users trade-off between the timeframe for result delivery and computational expenses. For the rest of the paper, the terms *user* and *end-users* will be used interchangeably.

A Grid scheduler, often called *resource broker*, acts as an interface between the user and distributed resources and hides the complexities of Grid computing [4][5]. It performs resource discovery, negotiates for access costs using trading services, maps jobs to resources (*scheduling*), stages the application and data for processing (*deployment*), starts job execution, and finally gathers the results. It is also responsible for monitoring and tracking application execution progress along with adapting to the changes in Grid runtime environment, variation in resource share availability, and failures.

In our Grid economy framework, the resource brokers use economy driven *deadline and budget constrained* (DBC) scheduling algorithms for allocating resources to application jobs in such a way that the users' requirements are met. In our early work [7], we developed three scheduling algorithms for cost, time, and time-variant optimisation strategies that support deadline and budget constraints. We implemented them within the Nimrod-G broker and explored their capability for scheduling task-farming or parameter-sweep and data-intensive computing applications such as drug design [12] on the WWG (World-Wide Grid) [9] testbed resources. To meet users' quality of service requirements, the broker leases Grid resources and services dynamically at runtime depending on their capability, cost, and availability.

In this work, we propose a new scheduling algorithm, called *DBC cost-time optimisation,* which extends the DBC cost-optimisation algorithm to optimise for time keeping the cost of computation at the minimum. Resources with the same cost are grouped together and time-optimisation scheduling strategy is applied while allocating jobs to a group. We demonstrate the ability of this new scheduling algorithm by implementing it within the economic Grid resource broker simulator built using the GridSim toolkit [3]. The performance of this new algorithm is evaluated by scheduling a synthetic task farming application on simulated WWG testbed resources for different deadline and budget scenarios. We then compare and contrast the results of scheduling with the cost optimisation algorithm.

The rest of this paper is organized as follows. Section 2 discusses the GridSim toolkit briefly and highlights the use of its features for simulating Grid environment and entities. The GridSim broker architecture and internal components that simulate and manage the execution of task farming applications along with scheduling algorithm are presented in Section 3. The simulation of heterogeneous resources with different capabilities and access costs, creation of synthetic application, and evaluation of proposed cost-time optimisation scheduling algorithm against the cost optimisation algorithm are discussed in Section 4. The final section summarizes the paper along with suggestions for future works.

## 2 GridSim: A Grid Modeling and Simulation Toolkit

The GridSim toolkit provides a comprehensive facility for simulation of different classes of heterogeneous resources, users, applications, resource brokers, and schedulers [3]. It has facilities for the modeling and simulation of resources and network connectivity with different capabilities, configurations, and domains. It supports primitives for application composition, information services for resource discovery, and interfaces for assigning application tasks to resources and managing their execution. These features can be used to simulate resource brokers or Grid schedulers for evaluating performance of scheduling algorithms or heuristics. We have used GridSim toolkit to create a resource broker that simulates Nimrod-G for design and evaluation of deadline and budget constrained scheduling algorithms with cost and time optimisations.

The GridSim toolkit resource modeling facilities are used to simulate the World-Wide Grid resources managed as time or space-shared scheduling policies. The broker and user entities extend the GridSim class to inherit ability for communication with other entities. In GridSim, application tasks/jobs are modeled as *Gridlet* objects that contain all the information related to the job and its execution management details such as job length in MI (Million Instructions), disk I/O operations, input and output file sizes, and the job originator. The broker uses GridSim's job management protocols and services to map a Gridlet to a resource and manage it throughout its lifecycle.

## 3 Simulation of Grid Resource Broker and Cost-Time Optimisation

The GridSim toolkit is used to simulate Grid environment and a Nimrod-G like deadline and budget constrained scheduling system called economic Grid resource broker. The simulated Grid environment contains multiple resources and user entities with different requirements. The user and broker entities extend the GridSim class. All the users create experiments that each of which contains application specification (a set of Gridlets that represent application jobs with different processing) and quality of service requirements (deadline and budget constraints with optimisation strategy). When simulation is started, the user entity creates an instance of its own broker entity and passes request for processing application jobs.

### 3.1.1 Broker Architecture

The broker entity architecture along with its interaction flow diagram with other entities is shown in Figure 1. The key components of the broker are: experiment interface, resource discovery and trading, scheduling flow manager backed with scheduling heuristics and algorithms, Gridlets dispatcher, and Gridlets receptor. A detailed discussion on the broker implementation using the GridSim toolkit can be found in [3]. However, to enable the understanding of the broker framework in which the new scheduling algorithm is implemented, we briefly present its operational model:

1. The user entity creates an experiment that contains application description (a list of Gridlets to be processed) and user requirements to the broker via the experiment interface.
2. The broker resource discovery and trading module interacts with the GridSim GIS entity to identify contact information of resources and then interacts with resources to establish their configuration and access cost. It creates a Broker Resource list that acts as placeholder for maintaining resource properties, a list of Gridlets committed for execution on the resource, and the resource performance data as predicted through the measurement and extrapolation methodology.
3. The scheduling flow manager selects an appropriate scheduling algorithm for mapping Gridlets to resources depending on the user requirements (deadline and budget limits; and optimisation strategy—cost, cost-time, time, or time variant). Gridlets that are mapped to a specific resource are added to the Gridlets list in the Broker Resource.
4. For each of the resources, the dispatcher selects the number of Gridlets that can be staged for execution according to the usage policy to avoid overloading resources with single user jobs.
5. The dispatcher then submits Gridlets to resources using the GridSim's asynchronous service.
6. When the Gridlet processing completes, the resource returns it to the broker's Gridlet receptor module, which then measures and updates the runtime parameter, *resource or MI share available to the user*. It aids in predicting the job consumption rate for making scheduling decisions.
7. The steps, 3–6, continue until all the Gridlets are processed or the broker exceeds deadline or budget limits. The broker then returns updated experiment data along with processed Gridlets back to the user entity.

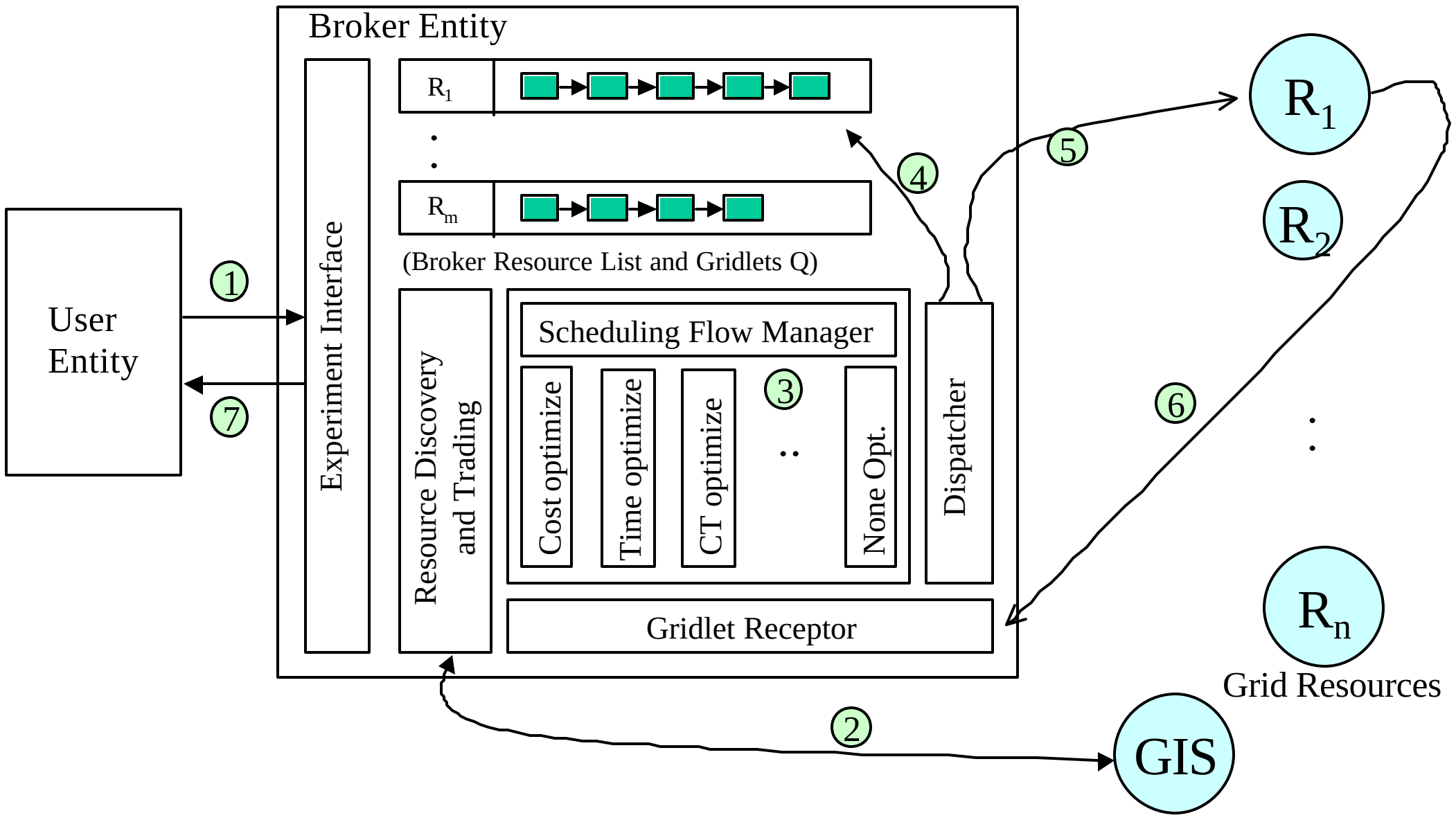

**Figure 1: Economic Grid resource broker architecture and its interaction with other entities.**

### 3.1.2 Deadline and Budget Constrained Cost-Time Optimisation Scheduling Algorithm

We have simulated deadline and budget constrained (DBC) scheduling algorithms, cost-optimisation [3], time-optimisation, and time-variant optimisation, presented in [7]. A new scheduling algorithm, called cost-time optimisation, proposed in this paper is shown in Figure 2. It extends the cost-optimisation algorithm to optimise the time without incurring additional processing expenses. This is accomplished by applying the time-optimisation algorithm to schedule jobs on resources having the same processing cost. The performance evaluation of this new algorithm is presented in the next section.

*Algorithm: DBC_Scheduling_with_Cost_Time_Optimisation()*

1. RESOURCE DISCOVERY: Identify the resources and their capability using the Grid information services.
2. RESOURCE TRADING: Identify *the cost* of all resources and *the capability* to be delivered per cost-unit. The *resource cost* can be expressed in units such as processing cost-per-MI, cost-per-job, CPU cost per time unit, etc. and the scheduler needs to choose suitable unit for comparison.
3. If the user supplies D and B-factors, then determine the absolute deadline and budget based on the capability of resources and their cost, and the application processing requirements (e.g., total MI required).
4. SCHEDULING: Repeat while there exists *unprocessed jobs* and the current time and processing expenses are within the deadline and budget limits. [It is triggered for each scheduling event or whenever a job completes. The event period is a function of deadline, job processing time, rescheduling overhead, resource share variation, etc.]:

   [SCHEDULE ADVISOR with Policy]

   a. For each resource, predict and establish the *job consumption rate* or *the available resource share* through the measure and extrapolation strategy taking into account the time taken to process previous jobs.
   b. SORT the resources by increasing order of *cost*. If two or more resources have the *same cost*, order them such that powerful ones (e.g., higher job consumption rate or resource share availability, but the first time based on the total theoretical capability, say the total MIPS) are preferred first.
   c. *Create resource groups* containing resources with the same cost.
   d. SORT the *resource groups* with the increasing order of cost.
   e. If any of the resource has jobs assigned to it in the previous scheduling event, but not dispatched to the resource for execution and there is variation in resource availability, then move appropriate number of jobs to the Unassigned-Jobs-List. This helps in updating the whole schedule based on the latest resource availability information.
   f. Repeat the following steps for each resource group as long as there exists unassigned jobs:
      i. *Repeat the following steps for each job in the Unassigned-Jobs-List depending on the processing cost and the budget availability:* [It uses the *time optimisation strategy*.]
      - Select a job from the Unassigned-Jobs-List.
      - For each resource, calculate/predict the job completion time taking into account previously assigned jobs and the job completion rate and resource share availability.
      - Sort resources by the increasing order of completion time.
      - Assign the job to the first resource and remove it from the Unassigned-Jobs-List if the predicted job completion time is less than the deadline.
5. [DISPATCHER with Policy]

   *Repeat the following steps for each resource if it has jobs to be dispatched:*
   - Identify the number of jobs that can be submitted without overloading the resource. Our default policy is to dispatch jobs as long as the number of user jobs deployed (active or in queue) is less than the number of PEs in the resource.

**Figure 2: Deadline and budget constrained (DBC) scheduling with cost-time optimisation.**

## 4 Scheduling Simulation and Performance Evaluation

To simulate application scheduling in GridSim environment using the economic Grid broker requires the modeling and creation of GridSim resources and applications that model jobs as Gridlets. In this section, we present resource and application modeling along with the results of scheduling experiments with quality of services driven application processing.

## 4.1 Resource Modeling

We modeled and simulated a number of time- and space-shared resources with different characteristics, configuration, and capability as those in the WWG testbed. We have selected the latest CPUs models AlphaServer ES40, Sun Netra 20, Intel VC820 (800EB MHz, Pentium III), and SGI Origin 3200 1X 500MHz R14k released by their manufacturers Compaq, Sun, Intel, and SGI respectively. The processing capability of these PEs in simulation time-unit is modeled after the base value of SPEC CPU (INT) 2000 benchmark ratings published in [10]. To enable the users to model and express their application processing requirements in terms of MI (million instructions) or MIPS (million instructions per second) on the standard machine, we *assume the MIPS rating of PEs is same as the SPEC rating*.

Table 1 shows the characteristics of resources simulated and their PE cost per time unit in G$ (Grid dollar). The simulated resources resemble the WWG testbed resources used in the Nimrod-G scheduling experiments reported in [11]. The access cost of PE in G$/time-unit not necessarily reflects the cost of processing when PEs have different capability. The brokers need to translate it into the G$ per MI for each resource. Such translation helps in identifying the relative cost of resources for processing Gridlets on them. It can be noted some of the resources in Table 1 have the same MIPS per G$. For example, both R4 and R8 have the same cost and so resources R2, R3, and R10.

**Table 1: World-Wide Grid testbed resources simulated using GridSim.**

| Resource Name in Simulation | Simulated Resource Characteristics<br>Vendor, Resource Type, Node OS, No of PEs | Equivalent Resource in Worldwide Grid<br>(Hostname, Location) | A PE SPEC/ MIPS Rating | Resource Manager Type | Price (G$/PE time unit) | MIPS per G$ |
|---|---|---|---|---|---|---|
| R0 | Compaq, AlphaServer, CPU, OSF1, 4 | grendel.vpac.org, VPAC, Melb, Australia | 515 | Time-shared | 8 | 64.37 |
| R1 | Sun, Ultra, Solaris, 4 | hpc420.hpcc.jp, AIST, Tokyo, Japan | 377 | Time-shared | 4 | 94.25 |
| R2 | Sun, Ultra, Solaris, 4 | hpc420-1.hpcc.jp, AIST, Tokyo, Japan | 377 | Time-shared | 3 | 125.66 |
| R3 | Sun, Ultra, Solaris, 2 | hpc420-2.hpcc.jp, AIST, Tokyo, Japan | 377 | Time-shared | 3 | 125.66 |
| R4 | Intel, Pentium/VC820, Linux, 2 | barbera.cnuce.cnr.it, CNR, Pisa, Italy | 380 | Time-shared | 1 | 380.0 |
| R5 | SGI, Origin 3200, IRIX, 6 | onyx1.zib.de, ZIB, Berlin, Germany | 410 | Time-shared | 5 | 82.0 |
| R6 | SGI, Origin 3200, IRIX, 16 | Onyx3.zib.de, ZIB, Berlin, Germany | 410 | Time-shared | 5 | 82.0 |
| R7 | SGI, Origin 3200, IRIX, 16 | mat.ruk.cuni.cz, Charles U., Prague, Czech Republic | 410 | Space-shared | 4 | 102.5 |
| R8 | Intel, Pentium/VC820, Linux, 2 | marge.csm.port.ac.uk, Portsmouth, UK | 380 | Time-shared | 1 | 380.0 |
| R9 | SGI, Origin 3200, IRIX, 4 (accessible) | green.cfs.ac.uk, Manchester, UK | 410 | Time-shared | 6 | 68.33 |
| R10 | Sun, Ultra, Solaris, 8, | pitcairn.mcs.anl.gov, ANL, Chicago, USA | 377 | Time-shared | 3 | 125.66 |

## 4.2 Application Modeling

We have modeled a task farming application that consists of 200 jobs. In GridSim, these jobs are packaged as Gridlets whose contents include the job length in MI, the size of job input and output data in bytes along with various other execution related parameters when they move between the broker and resources. The job length is expressed in terms of the time it takes to run on a standard resource PE with SPEC/MIPS rating of 100. Gridlets processing time is expressed in such a way that they are expected to take at least 100 time-units with a random variation of 0 to 10% on the positive side of the standard resource. That means, Gridlets' job length (processing requirements) can be at least 10,000 MI with a random variation of 0 to 10% on the positive side. This 0 to 10% random variation in Gridlets' job length is introduced to model heterogeneous tasks similar to those present in the real world parameter sweep applications.

## 4.3 DBC Scheduling Experiments with Cost and Cost-Optimisation Strategies

We performed both cost and cost-time optimisation scheduling experiments with different values of deadline and budget constraints (DBC) for a single user (*multiple users in the final paper*). The deadline is varied in simulation time from 100 to 3600 in steps of 500. The budget is varied from G$ 5000 to 22000 in steps of 1000. The number of Gridlets processed, deadline utilized, and budget spent for the DBC cost-optimisation scheduling strategy is shown in Figure 3a, Figure 3c, and Figure 3e, and for the cost-time optimisation scheduling strategy is shown in Figure 3b, Figure 3d, and Figure 3f. In both cases, when the deadline is low (e.g., 100 time unit), the number of Gridlets processed increases as the budget value increases. When a higher budget is available, the broker leases expensive resources to process more jobs within the deadline. Alternatively, when scheduling with a low budget value, the number of Gridlets processed increases as the deadline is relaxed.

The impact of budget for different values of deadline is shown in Figure 3e and Figure 3f for cost and cost-time strategies. For a larger deadline value (see the time utilization for deadline of 3600), the increase in budget value does not have much impact on resource selection. When the deadline is too tight (e.g., 100), it is likely that the complete budget is spent for processing Gridlets within the deadline.

It can be observed that the number of Gridlets processed and the budget-spending pattern is similar for both scheduling strategies. However, the time spent for the completion of all the jobs is significantly different (see Figure 3c and Figure 3d), as the deadline becomes relaxed. For deadline values from 100 to 1100, the completion time for both cases is similar, but as the deadline increases (e.g., 1600 to 3600), the experiment completion time for cost-time scheduling optimisation strategy is much less than the cost optimisation scheduling strategy. This is because when there are many resources with the same MIPS per G$, the cost-time optimisation scheduling strategy allocates jobs to them using the time-optimisation strategy for the entire deadline duration since there is no need to spent extra budget for doing so. This does not happen in case of cost-optimisation strategy—it allocates as many jobs that the first cheapest resource can complete by the deadline and then allocates the remaining jobs to the next cheapest resources.

A trace of resource selection and allocation using cost and cost-time optimisation scheduling strategies shown in Figure 4 indicates their impact on the application processing completion time. When the deadline is tight (e.g., 100), there is high demand for all the resources in short time, the impact of cost and cost-time scheduling strategies on the completion time is similar as all the resources are used up as long as budget is available to process all jobs within the deadline (see Figure 4a and Figure 4b). However, when the deadline is relaxed (e.g., 3100), it is likely that all jobs can be completed using the first few cheapest resources. In this experiment there were resources with the same cost and capability (e.g., R4 and R8), the cost optimisation strategy selected resource R4 to process all the jobs (see Figure 4c); whereas the cost-time optimisation strategy selected both R4 and R8 (see Figure 4d) since both resources cost the same price and completed the experiment earlier than the cost-optimisation scheduling (see Figure 3c and Figure 3d). This situation can be observed clearly in scheduling experiments with a large budget for different deadline values (see Figure 5). Note that the left most solid curve marked with the label "All" in the resources axis in Figure 5 represents the aggregation of all resources.

As the deadline increases, the cost optimisation algorithm predominantly scheduled jobs on the resource R4 (see Figure 5a) whereas, the cost-time optimisation algorithm scheduled jobs on resources R4 and R8 (see Figure 5a), the first two cheapest resources with same cost. Therefore, the application scheduling using the cost-time optimisation algorithm is able to finish earlier compared to the one scheduled using the cost optimisation algorithm (see Figure 6) and both strategies have spent the same amount of budget for processing its jobs (see Figure 7). The completion time for cost optimisation scheduling continued to increase with increase of the deadline as the broker allocated more jobs to the resource R4 and less to the resource R8. However, the completion time for deadline values 3100 and 3660 is the same as the previous one since the broker allocated jobs to only resource R4. This is not the case with the cost-time optimisation scheduling since jobs are allocated proportionally to both resources R4 and R8 and thus minimizing the completion time without spending any extra budget.

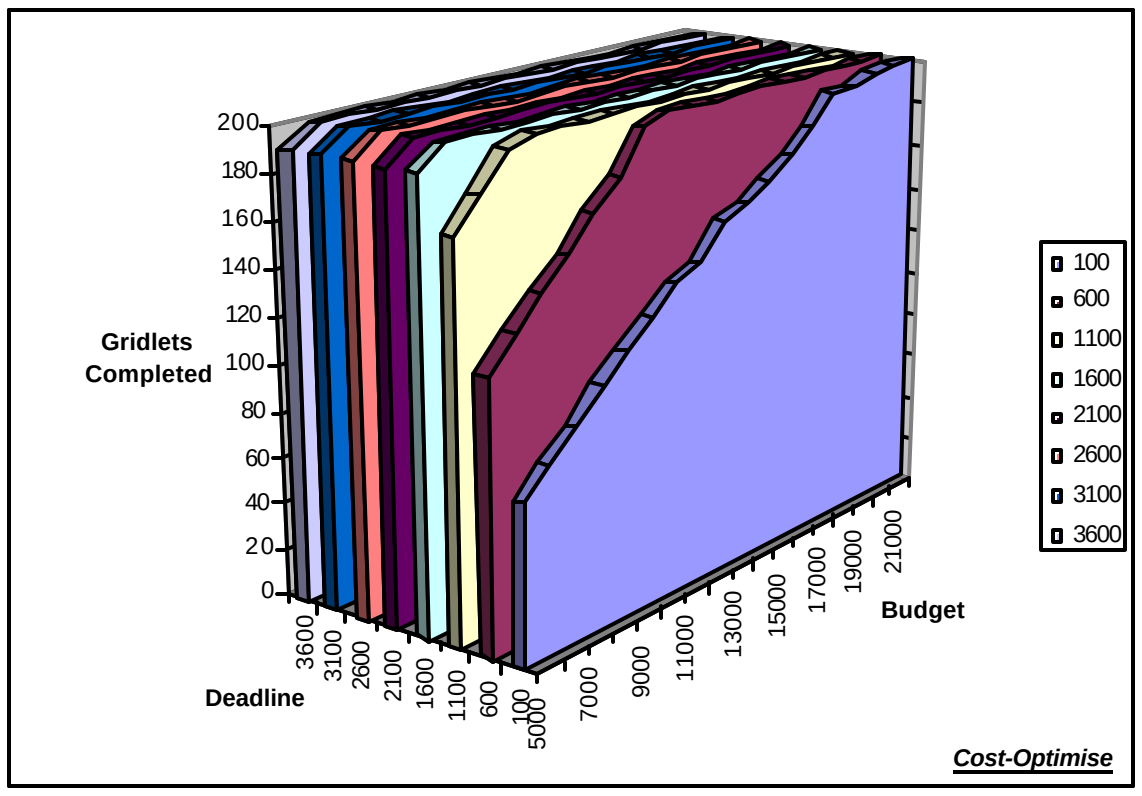

**(a) No. of Gridlets processed.**

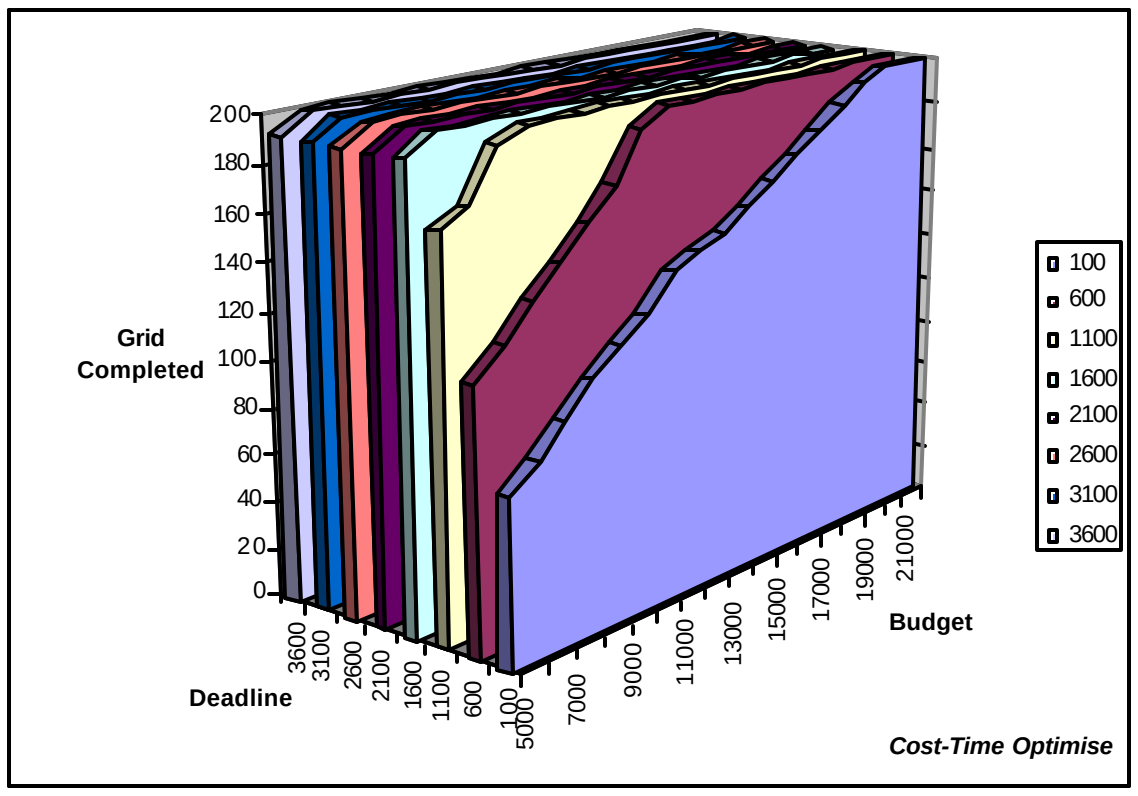

**(b) No. of Gridlets processed**

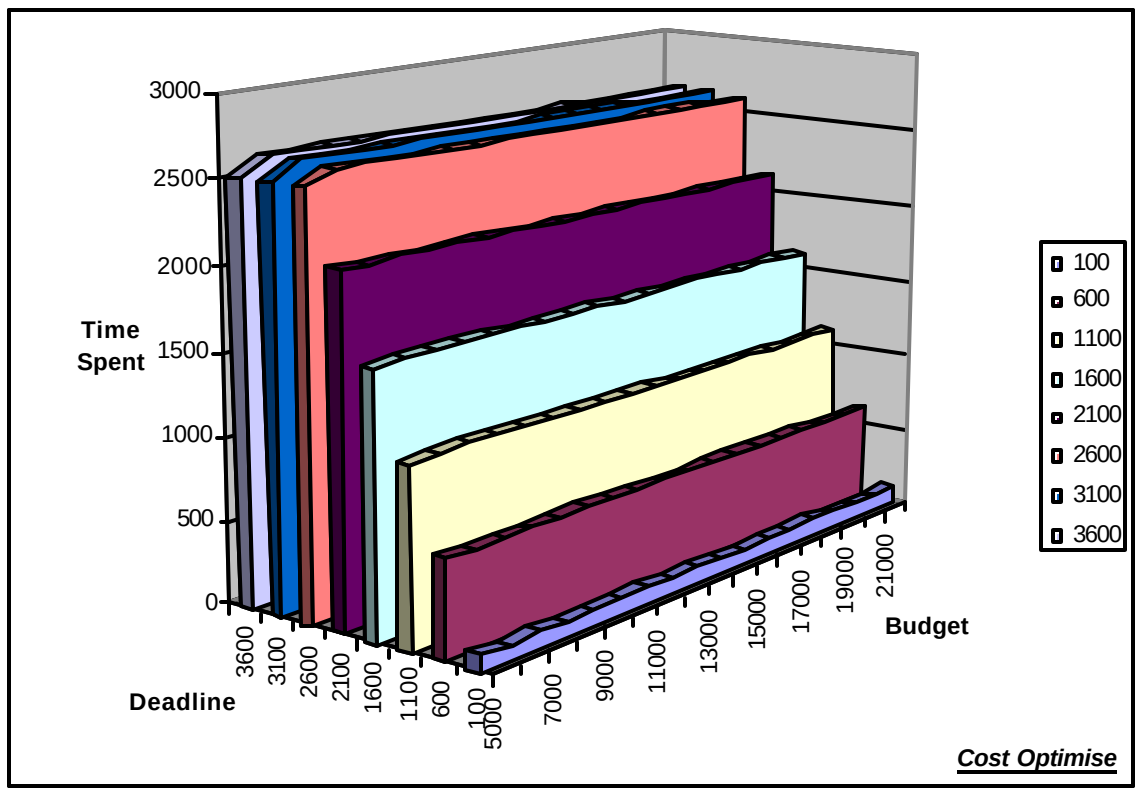

**(c) Time spent for processing Gridlets.**

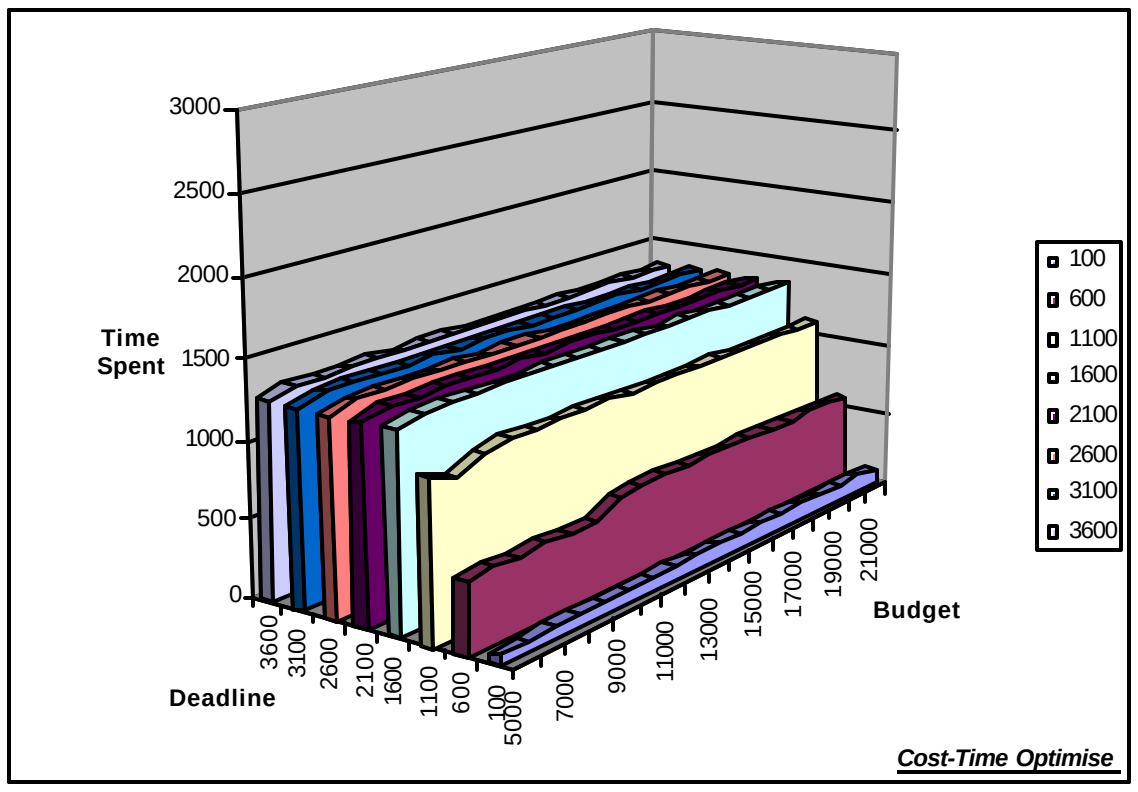

**(d) Time spent for processing Gridlets.**

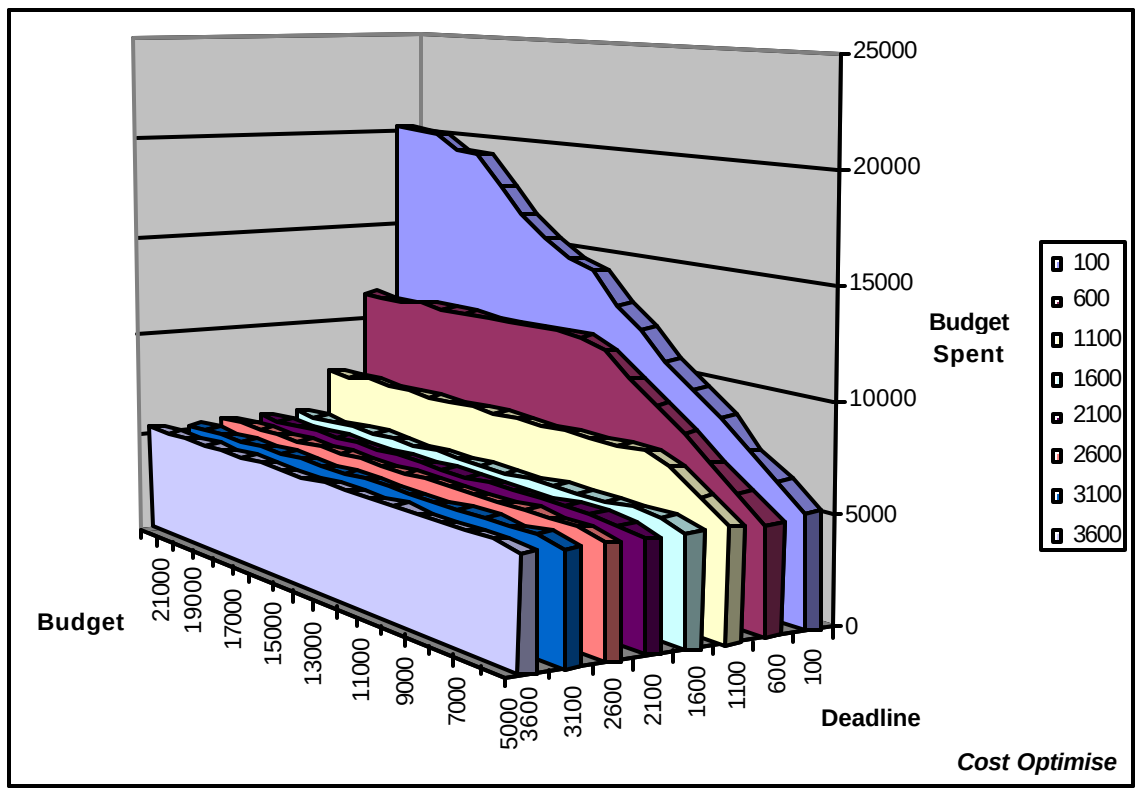

**(e) Budget spent for processing Gridlets.**

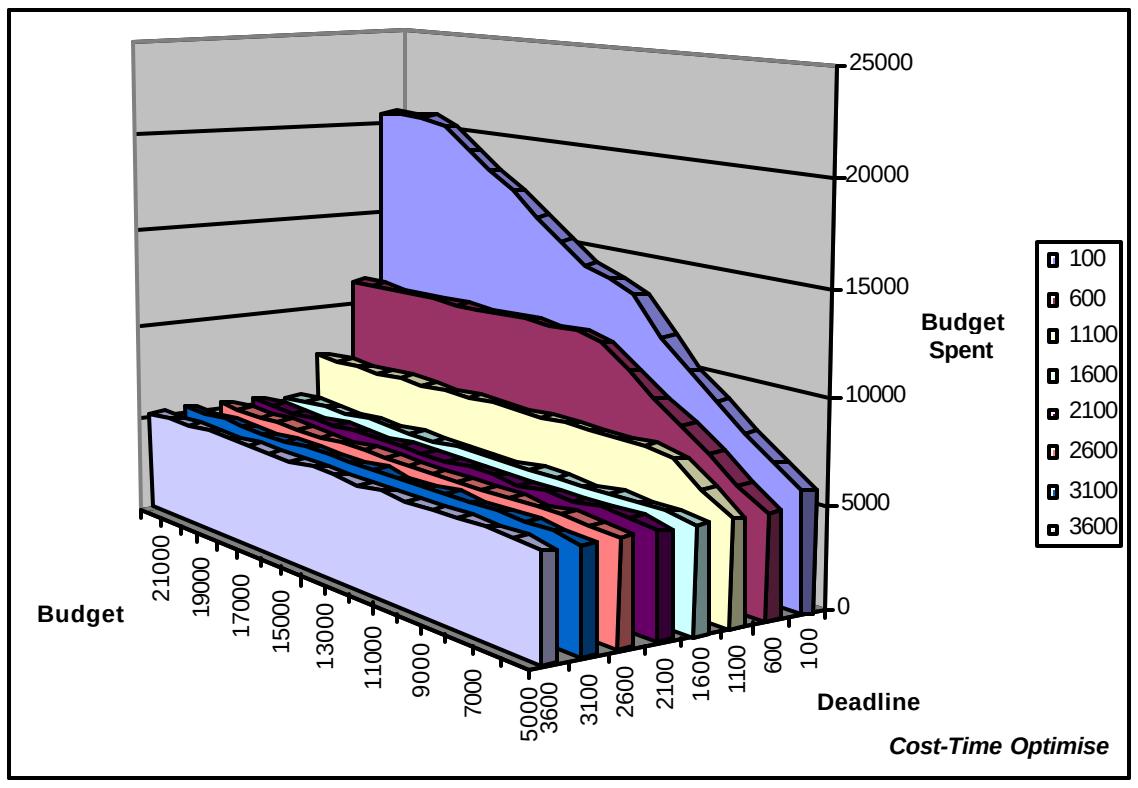

**(f) Budget spent for processing Gridlets.**

**Figure 3: The number of Gridlets processed, time, and budget spent for different deadline and time limits when scheduled using the cost and cost-time optimisation algorithms.**

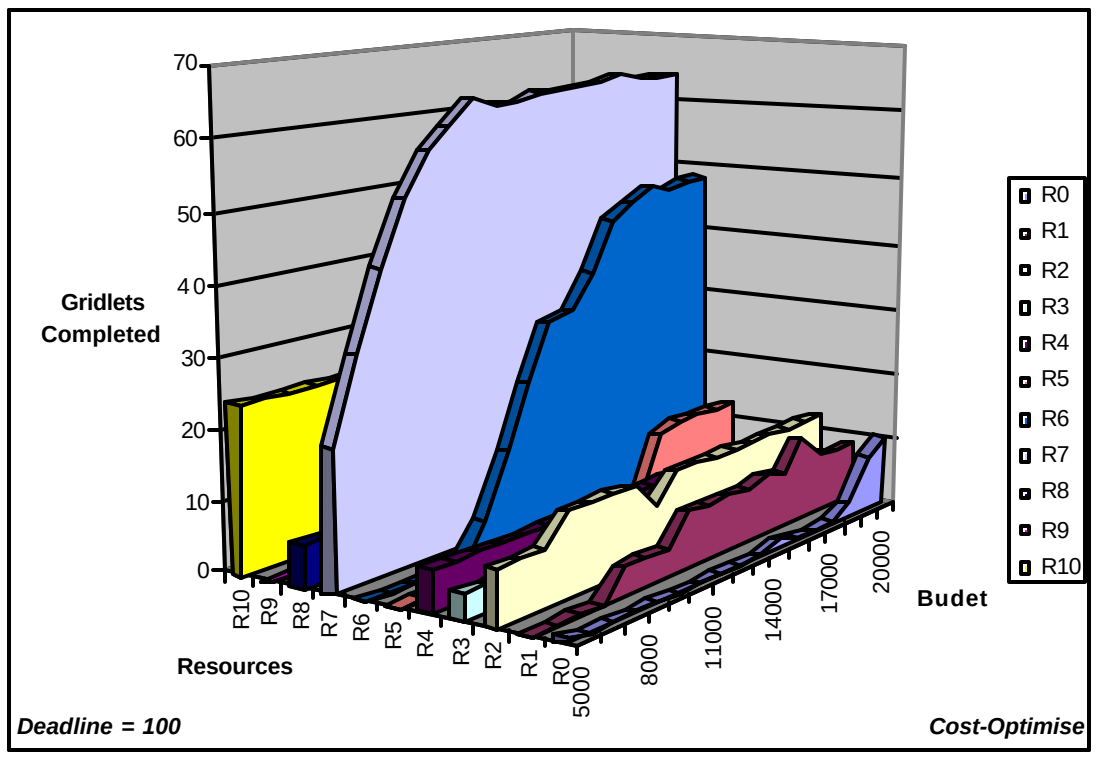

(a) Cost optimisation with a low deadline.

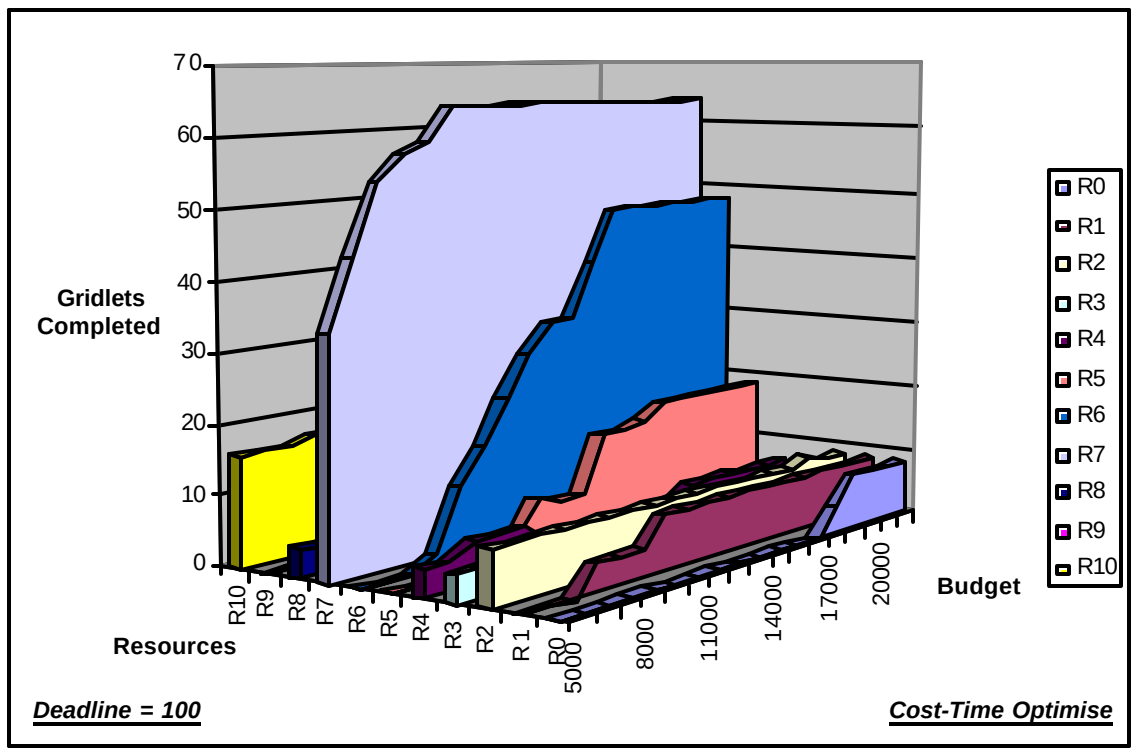

(b) Cost-time optimisation and a low deadline.

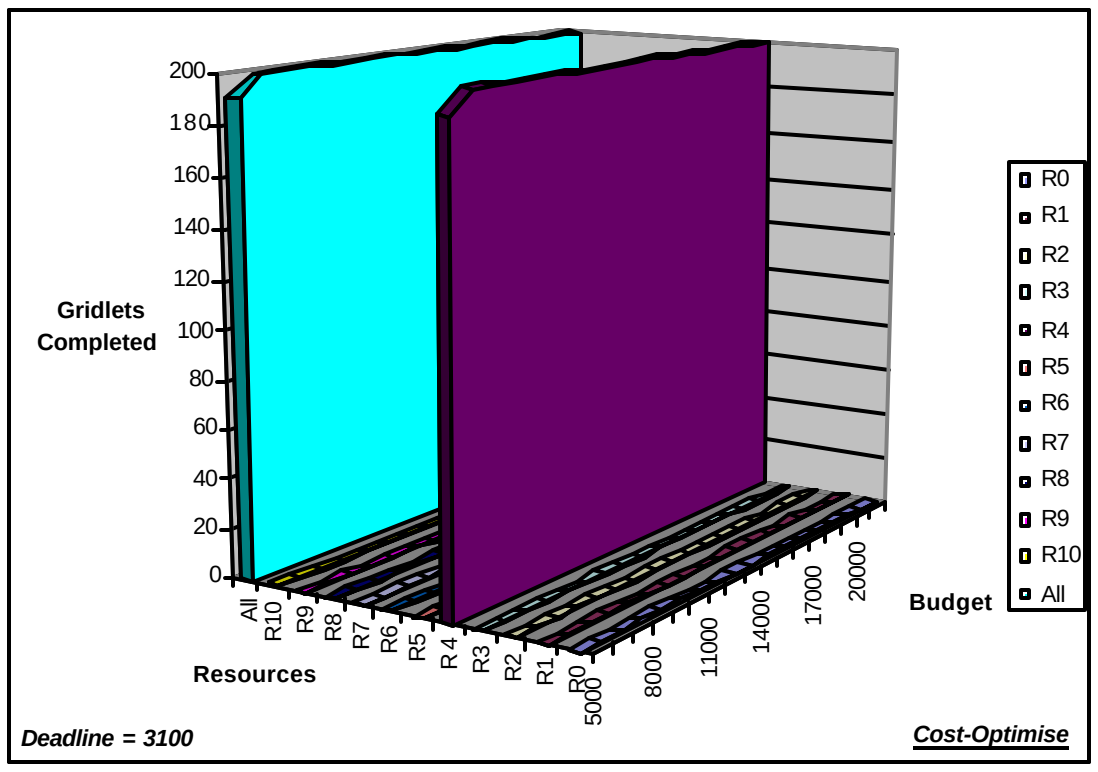

(c) Cost optimisation with a high deadline.

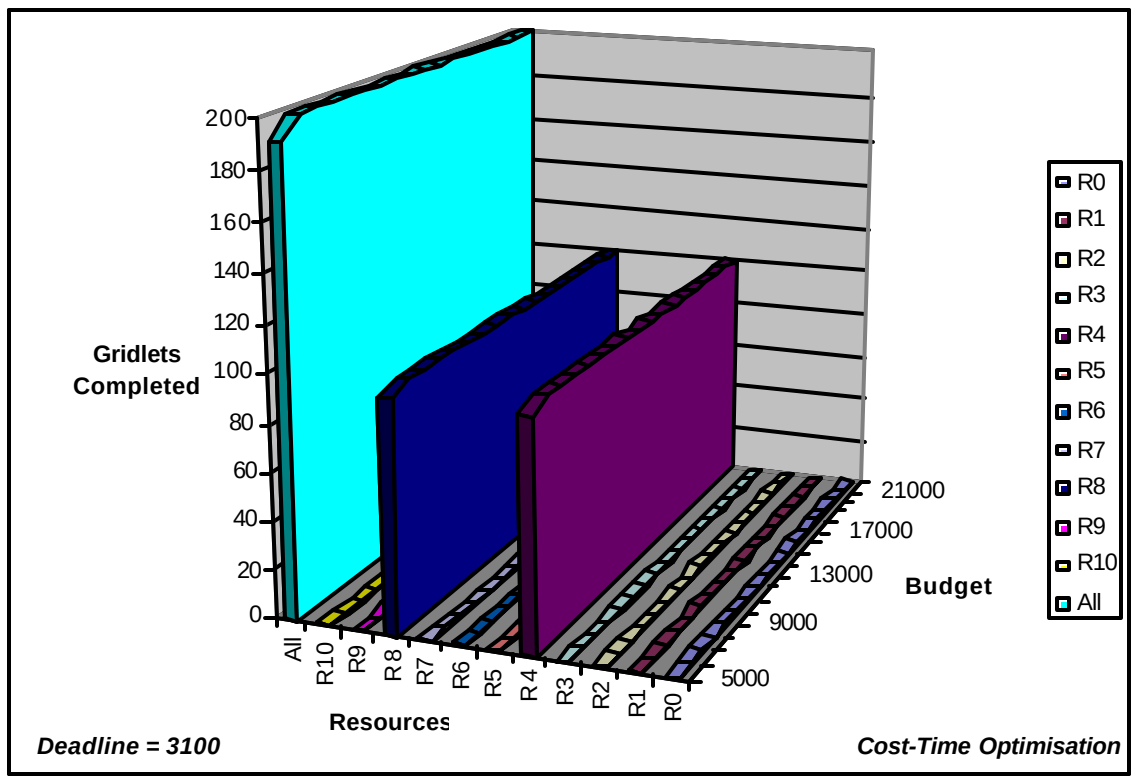

(d) Cost optimisati on with a high deadline.

**Figure 4: The number of Gridlets processed and resources selected for different budget values with a high deadline value when scheduled using the cost and cost-time optimisation algorithms.**

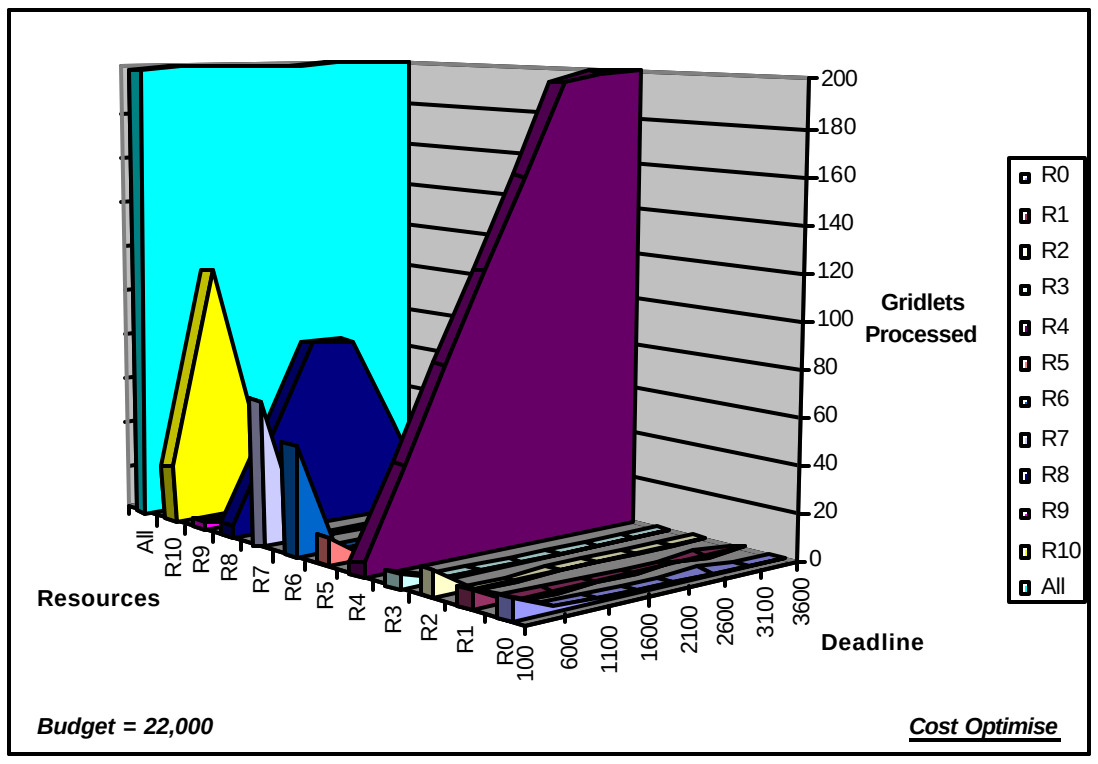

(a) Resource selection when the budget is high.

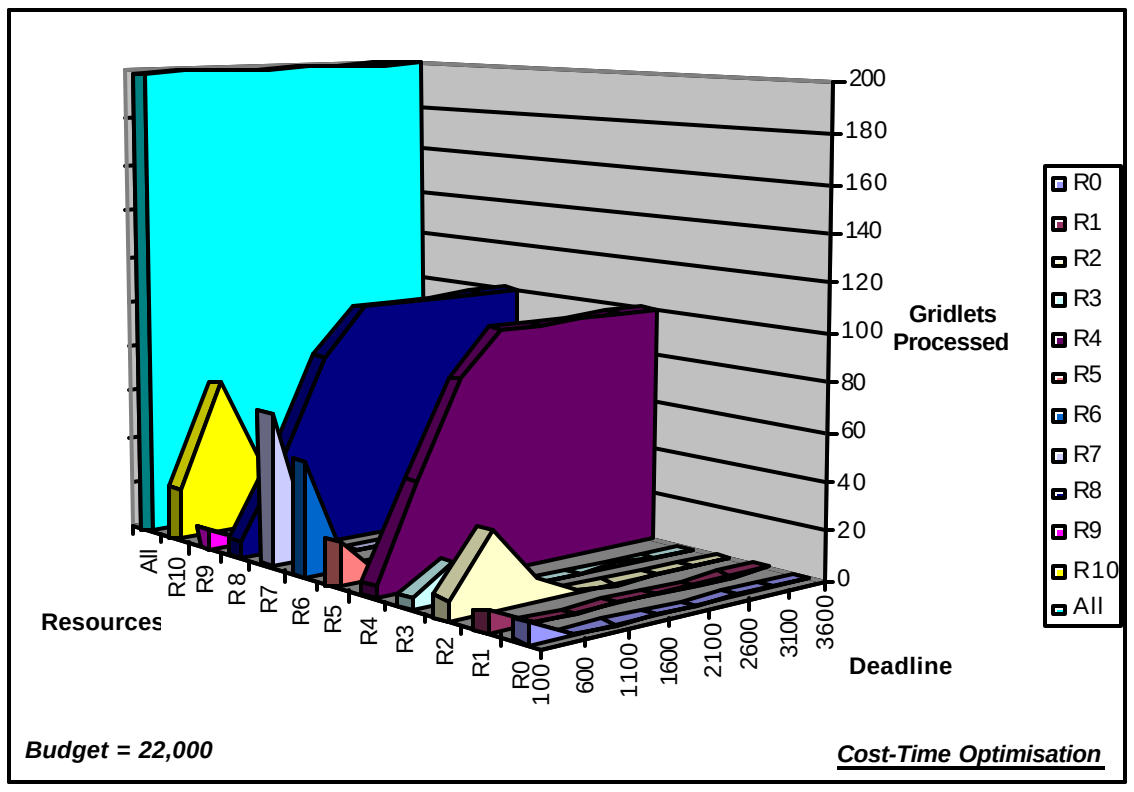

(b) Resource selection when the budget is high.

**Figure 5: The number of Gridlets processed and resources selected for different deadline values wi th a large budget when scheduled using the cost and cost-time optimisation algorithms.**

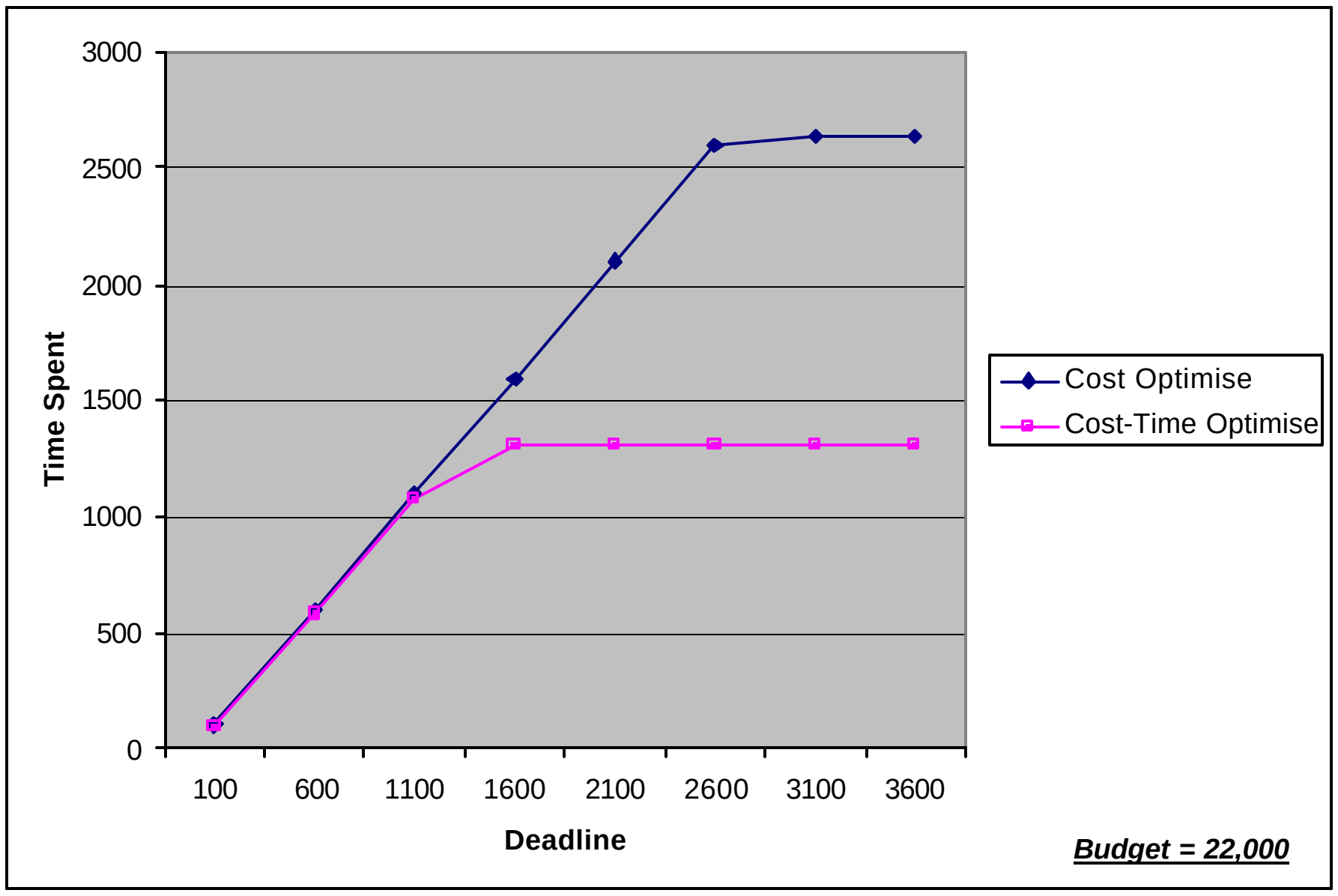

**Figure 6: The time spent for processing application jobs for different deadline constraints with a large budget when scheduled using the cost and cost-time optimisation algorithms.**

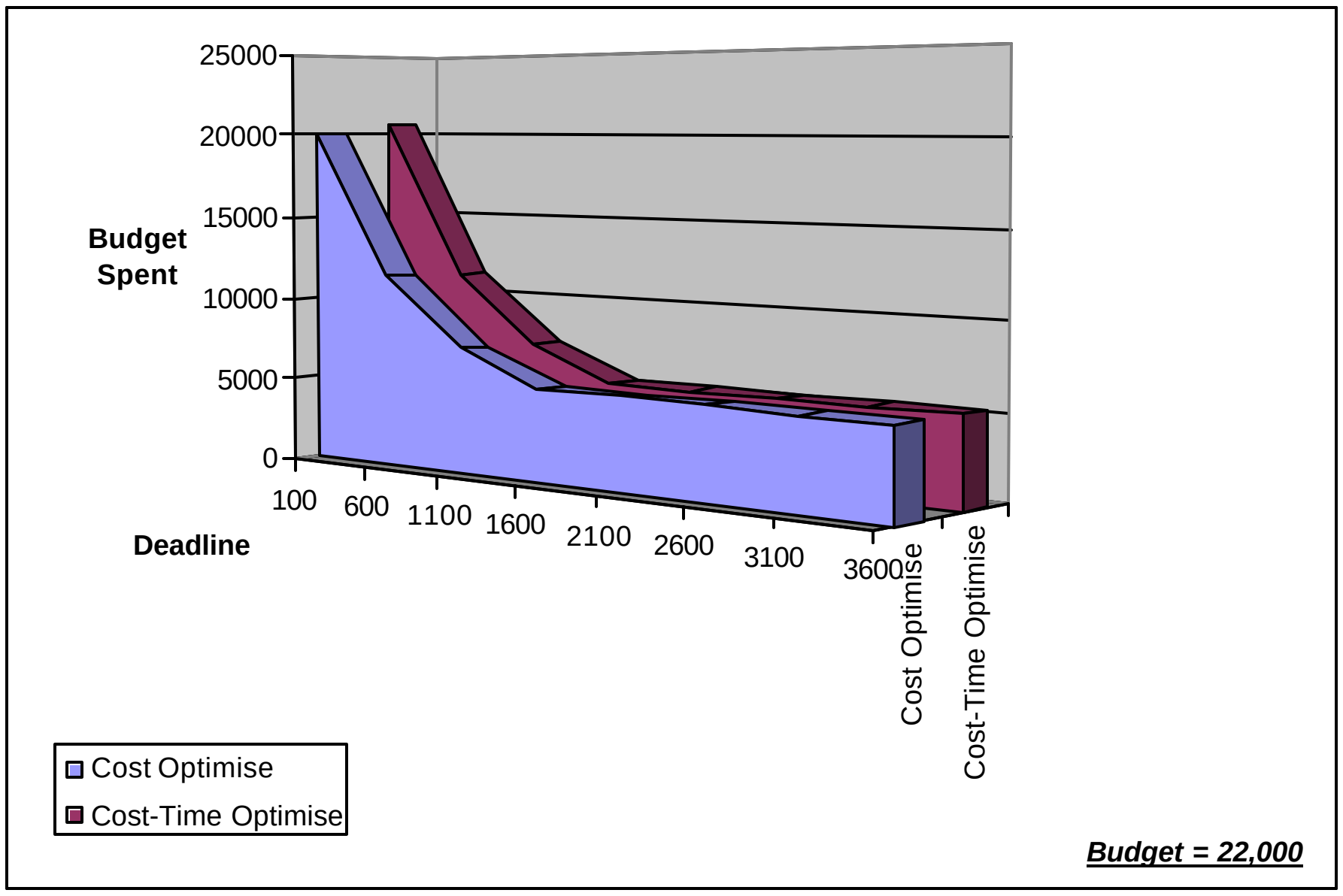

**Figure 7: The budget spent for processing application jobs for different deadline constraints with a large budget when scheduled using the cost and cost-time optimisation algorithms.**

Let us now take a microscopic look at the allocation of resources when a moderate deadline and large budget is assigned. A trace of resource allocation and the number of Gridlets processed at different times when scheduled using the cost and cost-time optimisation algorithms is shown in Figure 8 and Figure 9. It can be observed that for both the strategies, the broker used the first two cheapest resources, R4 and R8 fully. Since the deadline cannot be completed using only these resources, it used the next cheapest resources R2, R3, and R10 to make sure that deadline can be meet. The cost optimisation strategy allocated Gridlets to resource R10 only, whereas cost-time optimisation allocated Gridlets to resources R2, R3, and R10 as they cost the same price. Based on the availability of resources, the broker predicts the number of Gridlets that each resource can complete by the deadline and allocates to them accordingly (see Figure 10 and Figure 11). At each scheduling event, the broker evaluates the progress and resource availability and if there is any change, it reschedules some Gridlets to other resources to ensure that the deadline can be meet. This can be observed in Figure 10 and Figure 11—the broker allocated a few extra Gridlets to resource R10 (cost optimisation strategy) and resources R2, R3, and R10 (cost-time optimisation strategy) during the first few scheduling events.

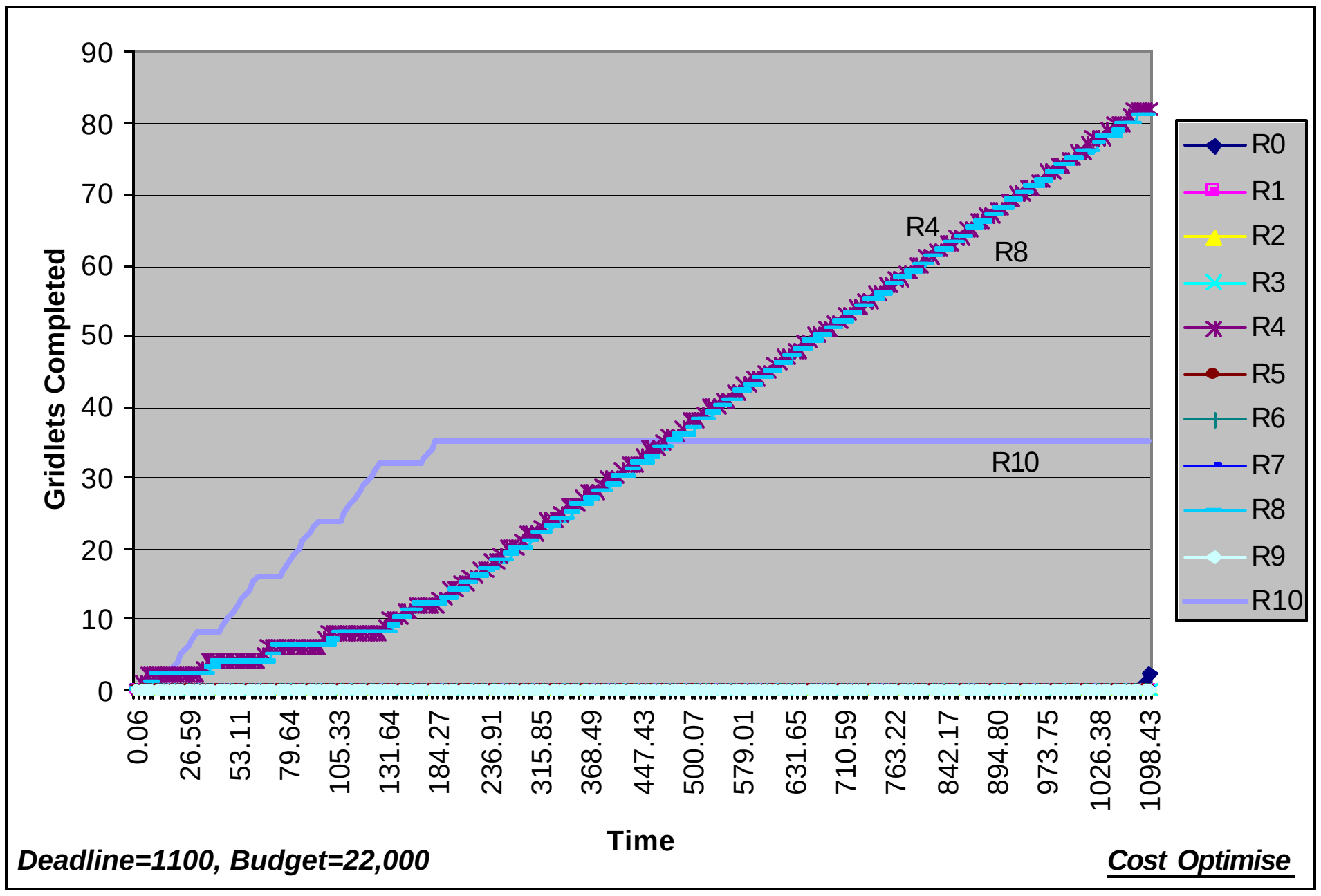

**Figure 8: Trace of No. of Gridlets processed on resources for a medium deadline and high budget constraints when scheduled using the cost optimisation strategy.**

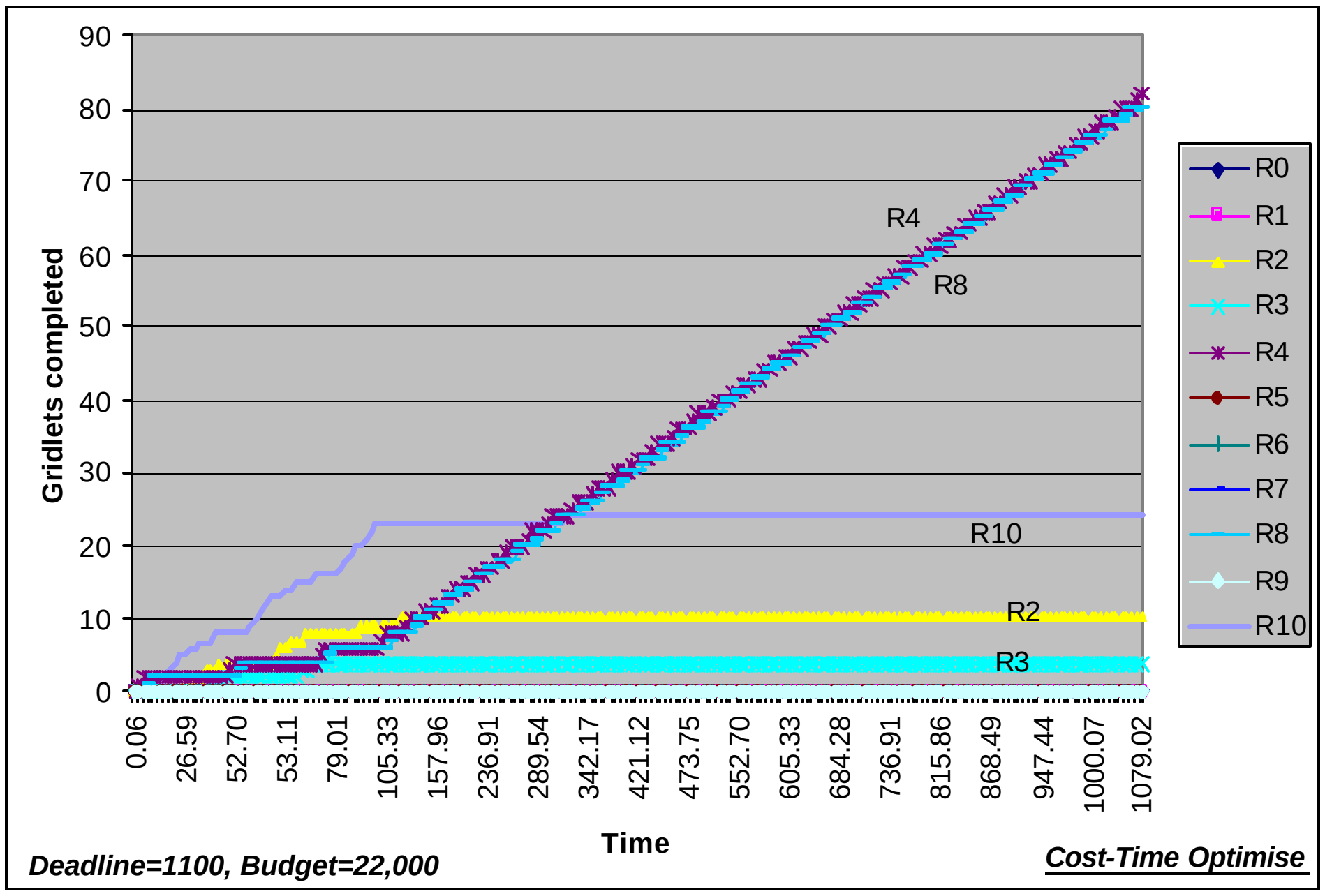

**Figure 9: Trace of No. of Gridlets processed on resources for a medium deadline and high budget constraints when scheduling using the cost-time optimisation strategy.**

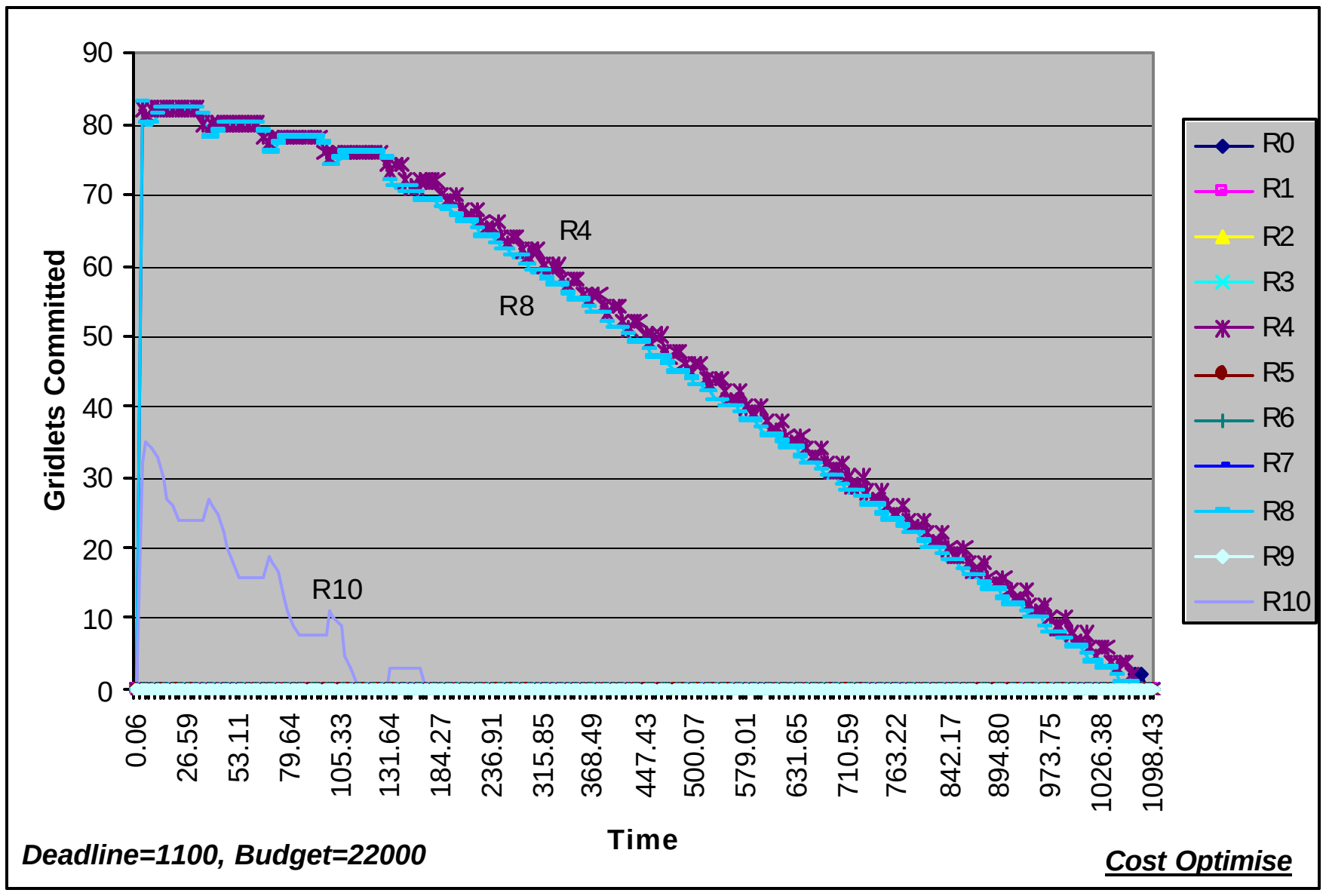

**Figure 10: Trace of the number of Gridlets committed to resources for a medium deadline and high budget constraints when scheduled using the cost optimisation strategy.**

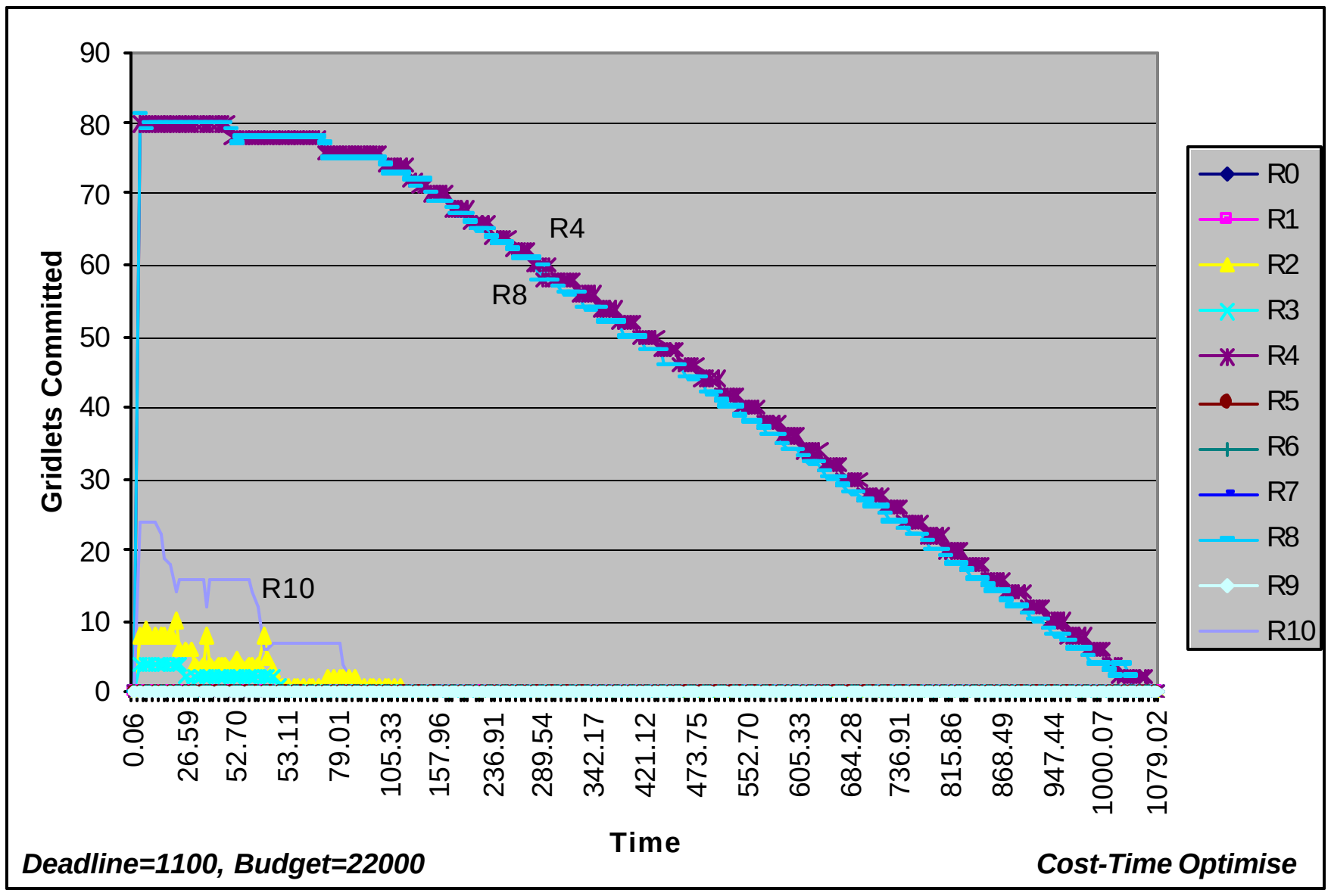

**Figure 11: Trace of the number of Gridlets committed to resources for a medium deadline and high budget constraints when scheduled using the cost-time optimisation strategy.**

## 5 Summary and Conclusion

Computational Grids enable the sharing, discovery, selection, and aggregation of geographically distributed heterogeneous resources for solving large-scale applications. We proposed computational economy as a metaphor for managing the complexity that is present in the management of distributed resources and allocation. It allows allocation of resources depending on the users' quality of service requirements such as the deadline, budget, and optimisation strategy. In this paper, we proposed a new deadline and budget constrained scheduling algorithm called *cost-time optimisation*. We developed a scheduling simulator using the GridSim toolkit and evaluated the new scheduling algorithm and compared its performance and quality of service delivery with cost optimisation algorithm. When there are multiple resources with the same cost and capability, the cost-time optimisation algorithm schedules jobs on them using the time-optimisation strategy for the deadline period. From the results of scheduling experiments for many scenarios with a different combination of the deadline and budget constraints, we observe that applications scheduled using the *cost-time* optimisation are able to complete earlier than those scheduled using the cost optimisation algorithm without incurring any extra expenses. This proves the superiority of the new deadline and budget constrained cost-time optimisation algorithm in scheduling jobs on global Grids.

Efforts are currently underway to implement the cost-time optimisation algorithm within the Nimrod-G Grid resource broker for scheduling parameter sweep applications on the World-Wide Grid testbed resources.

## Software Availability

The GridSim toolkit and the economic Grid broker simulator with source code can be downloaded from the GridSim project website:

`http://www.buyya.com/gridsim/`

## Acknowledgements

We thank Rob Gray (DSTC) for his comments on improving the paper and Marcelo Pasin (Federal University of Santa Maria, Brazil) for his help in modeling resources with SPEC benchmark ratings.